\documentclass[aps,prl,twocolumn,superscriptaddress]{revtex4-1}

\usepackage{graphicx,color}
\usepackage{epsfig}
\usepackage{dcolumn}
\usepackage{amsmath,amssymb,bm}
\usepackage{amsmath}
\usepackage{hyperref,url}
\usepackage[noabbrev]{cleveref}
\usepackage{CJK}
\usepackage{lineno}
\usepackage[english]{babel}
\usepackage[T1]{fontenc}
\usepackage{pifont}
\usepackage{amsfonts}
\usepackage{wasysym}
\usepackage{amsbsy}
\usepackage{psfrag}
\usepackage{color}
\usepackage{multirow}
\usepackage{cases}
\usepackage{stackengine}
\usepackage{float}
\usepackage{blkarray}
\usepackage{cancel}
\usepackage{mathrsfs}
\usepackage{empheq}
\usepackage{gensymb}
\usepackage{newtxtext}
\usepackage{newtxmath}
\usepackage[dvipsnames]{xcolor}

\addto\captionsenglish{}

\DeclareMathAlphabet{\textbfsf}{\encodingdefault}{\sfdefault}{bx}{sl}

\hypersetup{
    colorlinks=true,
    linkcolor=blue,
    citecolor=blue,
    urlcolor=blue,
}

\usepackage{fancyhdr}
\begin{document}



\title{Elastic turbulence homogenizes fluid transport in stratified porous media}



\author{Christopher A. Browne}
\affiliation{Department of Chemical and Biological Engineering, Princeton University, Princeton, NJ 08544, USA}

\author{Richard B. Huang}
\affiliation{Department of Chemical and Biological Engineering, Princeton University, Princeton, NJ 08544, USA}

\author{Callie W. Zheng}
\affiliation{Department of Chemical and Biological Engineering, Princeton University, Princeton, NJ 08544, USA}

\author{Sujit S. Datta}
\email{ssdatta@princeton.edu}
\affiliation{Department of Chemical and Biological Engineering, Princeton University, Princeton, NJ 08544, USA}



\begin{abstract}
Many key environmental, industrial, and energy processes rely on controlling fluid transport within subsurface porous media. These media are typically structurally heterogeneous, often with vertically-layered strata of distinct permeabilities---leading to uneven partitioning of flow across strata, which can be undesirable. Here, using direct in situ visualization, we demonstrate that polymer additives can homogenize this flow by inducing a purely-elastic flow instability that generates random spatiotemporal fluctuations and excess flow resistance in individual strata. In particular, we find that this instability arises at smaller imposed flow rates in higher-permeability strata, diverting flow towards lower-permeability strata and helping to homogenize the flow. Guided by the experiments, we develop a parallel-resistor model that quantitatively predicts the flow rate at which this homogenization is optimized for a given stratified medium. Thus, our work provides a new approach to homogenizing fluid and passive scalar transport in heterogeneous porous media.
\end{abstract}

\date{\today}

\maketitle

\section{Introduction}
\label{sec:intro}

Many key environmental, industrial, and energy processes---such as remediation of contaminated groundwater aquifers \citep{smith2008,hartmann2021risk}, recovery of oil from subsurface reservoirs \citep{durst1981,sorbie2013}, and extraction of heat from geothermal reservoirs \citep{di2021impact}---rely on the injection of a fluid into a subsurface porous medium. Such media are formed by sedimentary processes, often leading to vertically-layered strata of distinct pore sizes oriented along the direction of macroscopic flow \citep{freeze1975stochastic,dagan2012flow}. The permeability differences between these strata cause uneven fluid partitioning across them, with preferential flow through higher-permeability regions and ``bypassing'' of lower-permeability regions \citep{lake1981taylor, di2021impact}. This flow heterogeneity reduces the efficacy of contaminant remediation, oil recovery, and heat extraction from bypassed regions---necessitating the development of new ways to spatially homogenize the flow.

Low-molecular weight polymer additives have a long history of use in such applications to increase the injected fluid viscosity and thereby suppress instabilities, like viscous fingering, at immiscible (e.g., water-oil) interfaces \citep{durst1981,smith2008,sorbie2013}. However, this process of conformance control still suffers from the issue of uneven partioning of flow across different strata due to differences in permeability. Quantitatively, the superficial velocity in a given stratum $i$ is given by Darcy's law, representing each stratum as a homogeneous medium with uniformly-disordered pores of a single mean size: $U_{i}\equiv Q_{i}/A_{i}=(\Delta P/L)k_{i}/\eta_{\text{app}}$, where $Q_{i}$ is the volumetric flow rate through the stratum, $\Delta P$ is the pressure drop across a length $L$ of the parallel strata, $A_{i}$ and $k_{i}$ are the cross-sectional area and permeability of the stratum, respectively, and $\eta_{\text{app}}$ is the ``apparent viscosity'' of the polymer solution quantifying the macroscopic resistance to flow through the tortuous pore space. For low-molecular weight polymer additives, $\eta_{\text{app}}$ is simply given by the dynamic shear viscosity $\eta$ of the solution, and is typically not strongly dependent on flow rate or porous medium geometry. Therefore, differences in $k_{i}$ result in differences in $U_{i}$ between strata---leading to uneven partitioning of the flow across the entire stratified medium. 

Conversely, the apparent viscosity of a high-molecular weight polymer solution \textit{can} depend on flow rate. For many such solutions, $\eta_{\text{app}}$ strongly increases above a threshold flow rate in a homogeneous porous medium, even though $\eta$ of the bulk solution decreases with increasing shear rate \citep{marshall1967flow,james1975laminar,durst1981,dursthaas1981,ChauveteauMoan,kauser,hawardodell,odellhaward,zamani2015effect,clarke2016,skauge2018polymer,ibezim2021viscoelastic}. Direct visualization of the flow in a homogeneous medium \citep{Browne2021} recently established that this anomalous increase reflects the onset of a purely-elastic flow instability arising from the buildup of polymer elastic stresses during transport \citep{larson1990,shaqfeh1996,mckinley1996,pakdel1996,burghelea2004chaotic,rodd2007,afonso2010purely,zilz2012,galindo2012,ribeiro2014,clarke2016,machado2016extra,kawale2017a,qin2019flow,sousa2018purely,browne2019pore,Browne2020,walkama2020disorder,haward2021stagnation}. Specifically, this instability leads to ``elastic turbulence'', in which the flow exhibits random fluctuations reminiscent of inertial turbulence, despite the vanishingly small Reynolds numbers $\mathrm{Re}$ \citep{groisman2000,pan2013,qin2019flow,datta2021perspectives}---contributing added viscous dissipation that generates this anomalous increase in $\eta_{\text{app}}$ \citep{Browne2021}. In a stratified medium, this flow rate-dependence of $\eta_{\text{app},i}$ in each stratum may provide an avenue to break the proportionality between $k_{i}$ and $U_{i}$, potentially mitigating the uneven partitioning of the flow across strata. However, this possibility remains unexplored; indeed, it is still unknown how exactly elastic turbulence arises in each stratum.

Here, we demonstrate that elastic turbulence \textit{can} help homogenize flow in stratified porous media. Using pore-scale confocal microscopy and macro-scale imaging of passive scalar transport, we visualize the flow in a model porous medium with two distinct parallel strata, imposing a constant flow rate $Q$ through the entire medium. For small $Q$, the flow in both strata is laminar, leading to the typical uneven partitioning of flow across the strata. Strikingly, for $Q$ above a threshold value, elastic turbulence arises solely in the higher-permeability stratum and fluid is redirected to the lower-permeability stratum, helping to homogenize the flow. Above an even larger threshold flow rate, elastic turbulence also arises in this lower-permeability stratum, suppressing this flow redirection---leading to a window of flow rates at which this homogenization arises. Guided by these findings, we develop a parallel-resistor model that treats each stratum $i$ as a homogeneous medium with specified $A_i$, $k_i$, and therefore, $\eta_{\text{app},i}$, all coupled at the inlet and outlet. This model quantitatively captures the overall pressure drop across the stratified medium as well as the observed flow redirection with varying flow rate. It also elucidates the underlying cause of this redirection. In particular, above the first threshold flow rate, preferential flow causes elastic turbulence to arise solely in the higher-permeability stratum. The corresponding increase in the resistance to flow, as quantified by $\eta_{\text{app},i}$, redirects flow towards the lower-permeability stratum. Above the larger second threshold flow rate, the onset of elastic turbulence and corresponding increase in $\eta_{\text{app},i}$ in the lower-permeability stratum redirects flow back towards the higher-permeability stratum---yielding the experimentally-observed optimum in flow homogenization. Finally, we generalize this model, establishing the operating conditions at which this homogenization is optimized for porous media with arbitrarily many strata. Thus, our work provides a new approach to homogenizing fluid and passive scalar transport in heterogeneous porous media. Since many naturally-occurring media are stratified, we anticipate these findings to be broadly useful in environmental, industrial, and energy processes.

\section{Materials and Methods}
\label{sec:methods}
\begin{figure*}
    \centering
    \includegraphics[width=\textwidth]{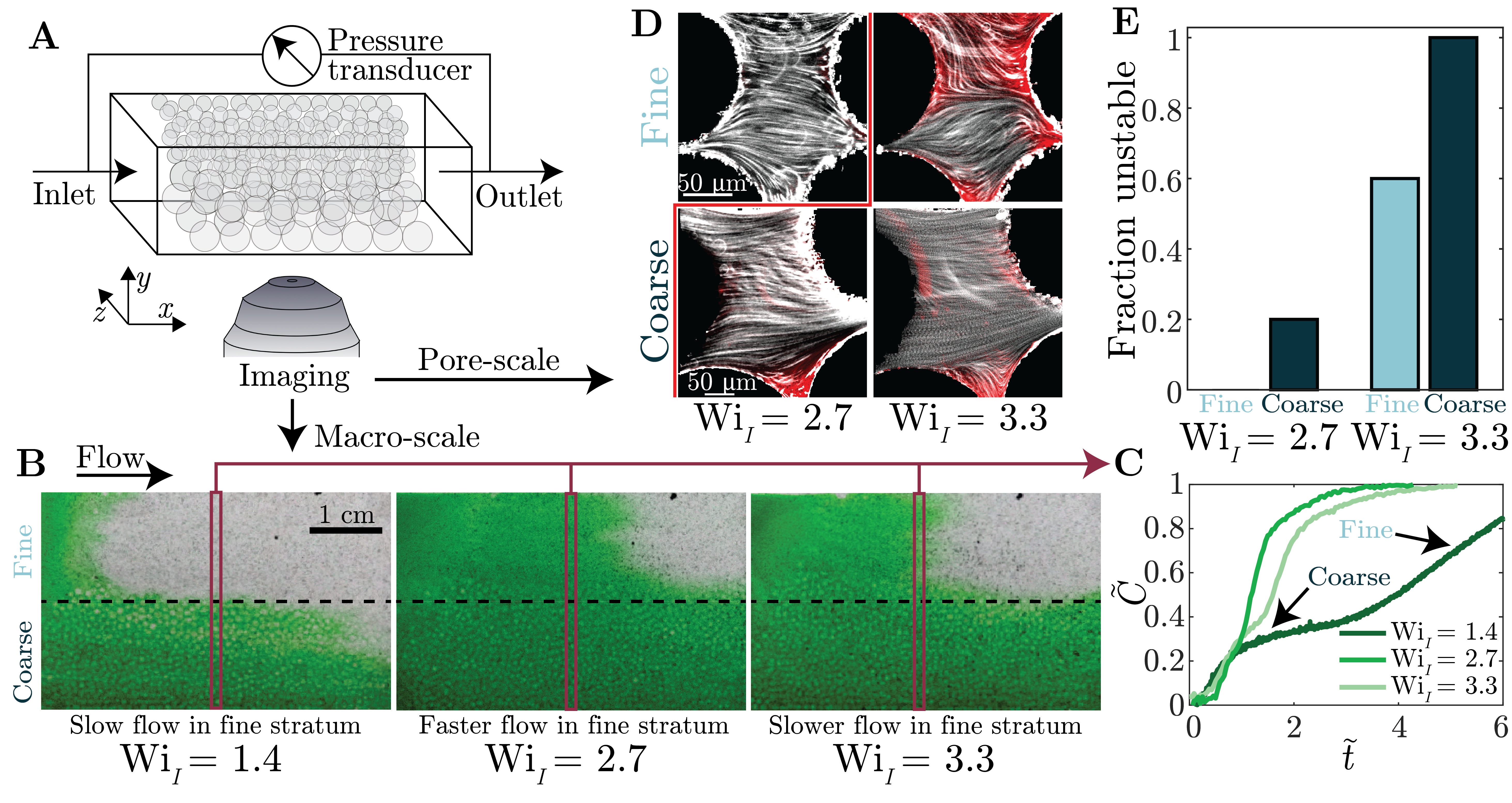}
    \caption{\textbf{Imaging reveals that an elastic polymer solution homogenizes the uneven flow between strata, coincident with the onset of elastic turbulence in the coarser stratum.} \textbf{A} Schematic of our model stratified porous media, with two parallel strata made of close-packed glass beads of distinct sizes. We characterize the flow using direct pore- or macro-scale flow visualization combined with pressure drop measurements across the medium. \textbf{B} Visualization of passive scalar transport by the polymer solution in a stratified Hele-Shaw assembly using a green dye. All images are taken at the same $\tilde{t}\equiv t/(0.5t_{PV})=2.5$, where time $t$ has been normalized by the time to fill half of the entire pore space volume. Due to the higher permeability of the coarse stratum (bottom), dye infiltrates faster than in the fine stratum (top). However, at the intermediate $\mathrm{Wi}_I=2.7$, this uneven partioning of the flow is reduced.  \textbf{C} Scalar breakthrough curves obtained by measuring the normalized dye concentration $\tilde{C}$ at the midpoint $x=L/2$ over time. Uneven flow partitioning at $\mathrm{Wi}_I=1.4$ leads to distinct jumps and prolongs $\tilde{C}$ to long times; by contrast, redirection of flow to the fine stratum at the intermediate $\mathrm{Wi}_I=2.7$ leads to more uniform and rapid breakthrough, shown by the smoother and earlier rise in $\tilde{C}(\tilde{t})$. This homogenization is mitigated at the even larger $\mathrm{Wi}_I=3.3$. \textbf{D} Streamline images of representative pores in a stratified microfluidic assembly; black circles are sections through the beads making up the solid matrix, white lines are time projections of the tracer particle pathlines that closely approximate the instantaneous flow streamlines. Imposed flow direction is from left to right. The flow homogenization at the intermediate $\mathrm{Wi}_I=2.7$ (first column) coincides with the onset of elastic turbulence solely in the coarse stratum (bottom)---indicated by the emergence of spatiotemporal fluctuations in the flow, shown by the red overlay whose intensity is given by the standard deviation in pixel intensity over the course of the time series of images. The mitigation of this homogenization at the even larger $\mathrm{Wi}_I=3.3$ (second column) coincides with the additional onset of elastic turbulence in the fine stratum, as well (top). \textbf{E} Fraction of 10 randomly-chosen pores observed in each stratum that exhibit unstable flow, defined as such by identifying whether fluid streamlines cross over the imaging duration. Only a small fraction of pores in the coarse stratum exhibit unstable flow at the intermediate $\mathrm{Wi}_I=2.7$, whereas a greater fraction of pores in both strata exhibit unstable flow at the larger $\mathrm{Wi}_I=3.3$---corroborating the results shown in \textbf{D}.}
    \label{fig:1}
\end{figure*}

To investigate the spatial distribution of flow in a stratified porous medium, we use imaging at two different length scales (Figure \ref{fig:1}A): macro-scale ($\sim100$s pores) and pore-scale ($\sim1$ pore). \\

\noindent\textbf{Macro-scale experiments in a Hele-Shaw assembly.} To characterize the macro-scale partitioning of flow, we fabricate an unconsolidated stratified porous medium in a Hele-Shaw assembly. We 3-D print an open-faced rectangular cell with span-wise ($y$-$z$-direction) cross-sectional area $A=3~\text{cm}\times0.4~\text{cm}$ and stream-wise ($x$-direction) length $L=5~\text{cm}$ using a clear methacrylate-based resin (FLGPCL04, \textit{Formlabs Form3}). To ensure an even distribution of flow at the boundaries, we use three inlets and outlets equally-spaced along the cross-section. We then fill the cell with spherical borosilicate glass beads of distinct diameters arranged in parallel strata using a temporary partition, with bead diameters $d_p=1000$ to 1400 $\muup$m (\textit{Sigma Aldrich}) and 212 to 255 $\muup$m (\textit{Mo-Sci}) for the higher-permeability coarse (subscript $C$) and lower-permeability fine (subscript $F$) strata, respectively. The strata have equal cross-sectional areas $A_C\approx A_F\approx A/2$ and thus their area ratio $\tilde{A}\equiv A_C/A_F\approx1$. Steel mesh with a 150 $\muup$m pore size cutoff placed over the inlet and outlet tubing prevents the beads from exiting the cell. We tamp down the beads for 30 min to form a dense random packing with a porosity $\phi_V\sim0.4$ \citep{onoda1990random}. We then screw the whole assembly shut with an overlying acrylic sheet cut to size, sandwiching a thin sheet of polydimethylsiloxane to provide a watertight seal. 

For all macro-scale experiments, we use a \textit{Harvard Apparatus} PHD 2000 syringe pump to first introduce the test fluid---either the polymer solution or the polymer-free solvent, which acts as a Newtonian control---at a constant flow rate $Q$ for at least the duration needed to fill the entire pore space volume $t_{PV}\equiv \phi_V AL/Q$ before imaging to ensure an equilibrated starting condition. We then visualize the macro-scale scalar transport by the fluid by introducing a step change in the concentration of a dilute dye (0.1 wt.\% green food coloring, \textit{McCormick}) and record the infiltration of the dye front using a DSLR camera (\textit{Sony} $\alpha$6300), as shown in figure \ref{fig:1}B. To track the progression of the dye as it is advected by the flow, we determine the ``breakthrough'' curve halfway along the length of the medium ($x=L/2$) by measuring the dye intensity $C$ averaged across the entire medium cross-section, normalized by the difference in intensities of the final dye-saturated and initial dye-free medium, $C_f$ and $C_0$, respectively: $\tilde{C}\equiv\left(\langle C\rangle_y-\langle C_0\rangle_y\right)/\left(\langle C_f\rangle_y-\langle C_0\rangle_y\right)$ (Figure \ref{fig:1}C). For all breakthrough curves thereby measured, time $t$ is normalized using the time taken to reach this halfway point, $\tilde{t}\equiv t/(0.5t_{PV})$. Repeating this procedure for individual strata (subscript $i$) and tracking the variation of the stream-wise position $X_i$ at which $\tilde{C}_i=0.5$ with time provides a measure of the superficial velocity $U_i=\mathrm{d}X_i/\mathrm{d}t$ in each stratum. {In between tests at different flow rates, we flush the assembly with the dye-free solution for at least ten pore volumes to remove any residual dye.}\\



\noindent\textbf{Pore-scale experiments in microfluidic assemblies.} To gain insight into the pore-scale physics, we use experiments in consolidated microfluidic assemblies. We pack spherical borosilicate glass beads (\textit{Mo-Sci}) in square quartz capillaries ($A=$ 3.2 mm $\times$ 3.2 mm{; \textit{Vitrocom}}), densify them by tapping, and lightly sinter the beads---resulting in dense random packings again with $\phi_V\sim0.4$ \citep{krummel2013visualizing}. We use this protocol to fabricate three different microfluidic media: a  homogeneous higher-permeability coarse medium ($d_p=$ 300 to 355 $\muup$m), a homogeneous lower-permeability fine medium ($d_p=$ 125 to 155 $\muup$m), and a stratified medium with parallel higher-permeability coarse and lower-permeability fine strata, each composed of the same beads used to make the homogeneous media, again with equal cross-section areas, $\tilde{A}\approx 1$ \citep{datta2013drainage, lu2020forced}. 
We measure the fully-developed pressure drop $\Delta P$ across each medium using an \textit{Omega PX26} differential pressure transducer. 

For all pore-scale experiments, before each experiment, {we infiltrate the medium to be studied first with isopropyl alcohol (IPA) to prevent trapping of air bubbles and then displace the IPA by flushing with water}. We then displace the water with the miscible polymer solution, seeded with 5 ppm of fluorescent carboxylated polystyrene tracer particles (\textit{Invitrogen}), $D_t=200~\mathrm{nm}$ in diameter. This solution is injected into the medium at a constant volumetric flow rate $Q$ {using \textit{Harvard Apparatus} syringe pumps---a PHD 2000 for $Q>1~\text{mL/hr}$ or a Pico Elite for $Q<1~\text{mL/hr}$---}for at least 3 hours to reach an equilibrated state before flow characterization. After each subsequent change in $Q$, the flow is given 1 hour to equilibrate before imaging. We monitor the flow in individual pores using a \textit{Nikon} A1R+ laser scanning confocal fluorescence microscope with a 488 nm excitation laser and a 500-550 nm sensor detector; the tracer particles have excitation between 480 and 510 nm with an excitation peak at 505 nm, and emission between 505 and 540 nm with an emission peak at 515 nm. These particles are faithful tracers of the underlying flow field since the P\'eclet number $\mathrm{Pe}\equiv (Q/A)D_t/\mathcal{D}>10^5\gg1$, where  $\mathcal{D}=k_BT/(3\pi\eta D_t)=6\times10^{-3}~\muup\mathrm{m^2/s}$ is the Stokes-Einstein particle diffusivity. We then visualize the flow using a 10$\times$ objective lens with the confocal resonant scanner, obtaining successive 8 $\muup\mathrm{m}$-thick optical slices at a $z$ depth {$\sim100$s $\muup\mathrm{m}$} within the medium. Our imaging probes an $x$-$y$ field of view $159~\muup\mathrm{m}\times159~\muup\mathrm{m}$ at 60 frames per second for pores with $d_p=125$ to $155~\muup\mathrm{m}$ or $318~\muup\mathrm{m}\times318~\muup\mathrm{m}$ at 30 frames per second for pores with $d_p=300$ to $355~\muup\mathrm{m}$. 

To monitor the flow in the different pores over time, we use an ``intermittent'' imaging protocol. Specifically, we record the flow in multiple pores chosen randomly throughout each medium (19 and 20 pores of the homogeneous coarse and fine media, respectively) for 2~s-long intervals every 4~min over the course of 1~h. For the experiments in homogeneous fine and stratified media, we also complement this protocol with ``continuous'' imaging in which we monitor the flow successively in 10 pores of the homogeneous fine medium for 5 min-long intervals each. For ease of visualization, we intensity-average the successive images thereby obtained over a time scale $\approx2.5~\mathrm{\muup m}/\left(Q/A\right)$ (Figure \ref{fig:1}D), producing movies of the tracer particle pathlines that closely approximate the instantaneous flow streamlines. \\

\noindent\textbf{Permeability measurements.} For each medium, we determine the permeability via Darcy's law using experiments with pure water. For the microfluidic assemblies, we obtain $k_C=79~\mathrm{\muup m^2}$ and $k_F=8.6~\mathrm{\muup m^2}$ for the homogeneous coarse and fine media, respectively---comparable to our previously-measured values on similar media \citep{krummel2013visualizing} and to the prediction of the established Kozeny-Carman relation \citep{philipse1993liquid}. The permeability ratio between the two strata is then $\tilde{k}\equiv k_C/k_F\approx9$. The measured permeability for the entire stratified porous medium is $k=32~\muup\mathrm{m}^2$, in reasonable agreement with the prediction obtained by considering the strata as separated homogeneous media providing parallel resistance to flow, $k\approx\tilde{A} k_C+(1-\tilde{A})k_F\approx44~\muup\mathrm{m}^2$. 

The permeability of an isolated stratum in a stratified medium varies as $\sim d_p^2$, similar to a homogeneous porous medium. Hence, for the Hele-Shaw assembly, we estimate the permeability of each stratum by scaling $k_C$ and $k_F$  with the differences in bead size. We thereby estimate $k\approx440~\muup\mathrm{m}^2$ ($\tilde{k}\approx26$) for the entire stratified medium, {in reasonable agreement with the measured $k=270~\muup\text{m}^2$}. 

For both assemblies, we define a characteristic shear rate of the entire medium $\dot{\gamma}_I\equiv Q/\left(A\sqrt{\phi_V k}\right)$ as the ratio between the characteristic pore flow speed $Q/(\phi_V A)$ and length scale $\sqrt{k/\phi_V}$ \citep{zami2016transition,berg2017shear}. Our experiments explore the range $\dot{\gamma}_I\approx0.2$ to $26~\mathrm{s}^{-1}$.\\

\begin{figure*}
    \centering
    \includegraphics[width=\textwidth]{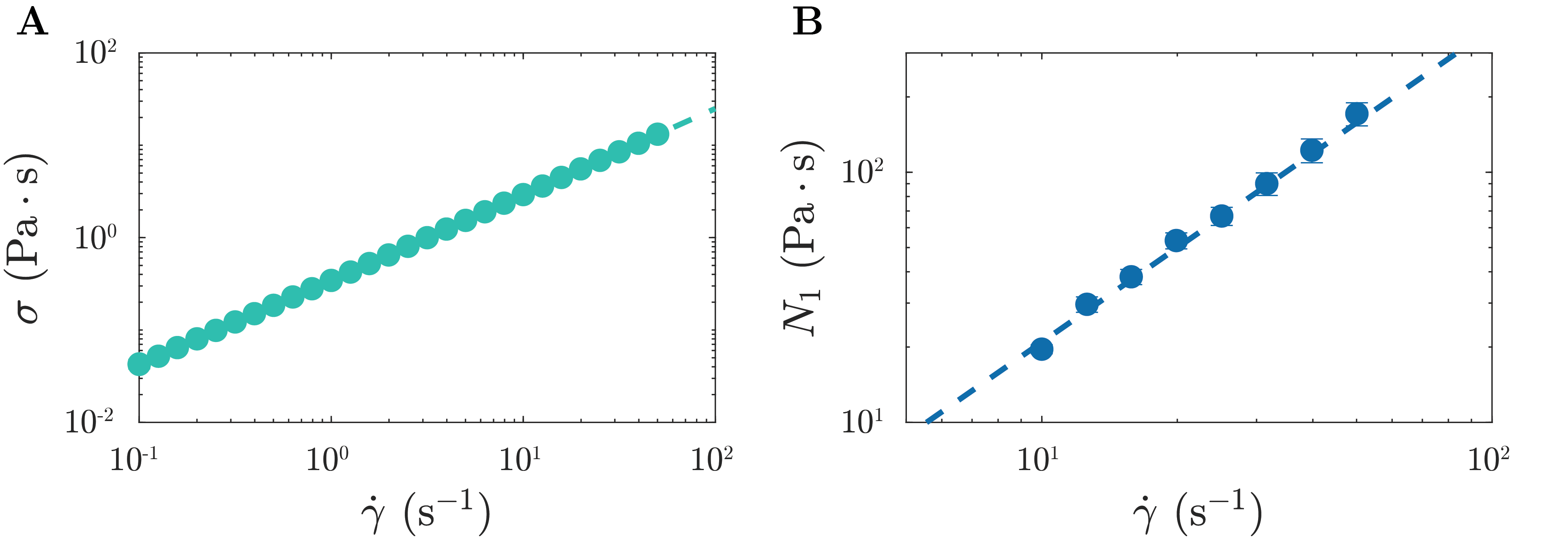}
    \caption{\textbf{Shear rheology of the bulk polymer solution}. \textbf{A} Shear stress $\sigma$ varies nearly linearly with the shear rate $\dot{\gamma}$, indicating that the solution approximates a Boger fluid; the dashed line shows the power-law fit $\sigma=A_s\dot{\gamma}^{\alpha_s}$ with $A_s=0.3428\pm0.0002~\mathrm{Pa\cdot s}^{\alpha_s}$ and $\alpha_s=0.931\pm0.001$. \textbf{B} First normal stress difference $N_1$ also increases with increasing shear rate $\dot{\gamma}$ ; the dashed line shows the power-law fit $N_1=A_n\dot{\gamma}^{\alpha_n}$ with $A_n=1.16\pm0.03~\mathrm{Pa\cdot s}^{\alpha_n}$ and $\alpha_n=1.25\pm0.02$. Error bars represent one standard deviation of multiple measurements; where they are not visible, they are smaller than the marker size.}
    \label{fig:rheology}
\end{figure*}

\noindent\textbf{Polymer solution rheology.} The polymer solution is a Boger fluid comprised of dilute 300 ppm 18 MDa partially hydrolyzed polyacrylamide (HPAM) dissolved in a viscous aqueous solvent composed of 6 wt.\% ultrapure \textit{milliPore} water, 82.6 wt.\% glycerol (\textit{Sigma Aldrich}), 10.4 wt.\% dimethylsulfoxide (\textit{Sigma Aldrich}),  and 1 wt.\% NaCl. This solution is formulated to precisely match its refractive index to that of the glass beads---thus rendering each medium transparent when saturated. From intrinsic viscosity measurements the overlap concentration is $c^{*}\approx0.77/[\eta]=600\pm300~\mathrm{ppm}$ \citep{Browne2021} and the radius of gyration is $R_g\approx220~\mathrm{nm}$ \citep{rubinstein2003polymer}, and therefore, our experiments use a dilute polymer solution at $\approx0.5$ times the overlap concentration. The shear stress $\sigma\left(\dot{\gamma}_I\right)=A_s\dot{\gamma}^{\alpha_s}$ and first normal stress difference $N_1\left(\dot{\gamma}_I\right)=A_n\dot{\gamma}^{\alpha_n}$ are measured in an \textit{Anton Paar} MCR301 rheometer, using a 1\degree~5 cm-diameter conical geometry set at a 50 $\muup$m gap, yielding the best-fit power laws $A_s=0.3428\pm0.0002~\mathrm{Pa\cdot s}^{\alpha_s}$ with $\alpha_s=0.931\pm0.001$ and $A_n=1.16\pm0.03~\mathrm{Pa\cdot s}^{\alpha_n}$ with $\alpha_n=1.25\pm0.02$ {(Figure \ref{fig:rheology})}. 

These measurements then enable us to calculate the characteristic interstitial Weissenberg number, which characterizes the role of polymer elasticity in the flow by comparing the magnitude of elastic and viscous stresses, $\mathrm{Wi}_I\equiv N_1\left(\dot{\gamma}_I\right)/\left(2\sigma\left(\dot{\gamma}_I\right)\right)$, as commonly defined \citep{datta2021perspectives}. In our experiments this quantity exceeds unity, ranging from 1 to 5.5, suggesting that viscoelastic flow instabilities likely arise in the flow \citep{larson1990,shaqfeh1996,pakdel1996, rodd2007,afonso2010purely,zilz2012,galindo2012,pan2013,ribeiro2014,sousa2018purely,browne2019pore,Browne2020,Browne2021}---as we directly verify using flow visualization, detailed further below. We also characterize the role of inertia with the Reynolds number $\mathrm{Re}=\rho Ud_p/\eta\left(\dot{\gamma}_I\right)$, which quantifies the ratio of inertial to viscous stresses for a fluid with density $\rho$. In our experiments this quantity ranges from $\mathrm{Re}=2\times10^{-7}$ to $2\times10^{-5}\ll1$, indicating that inertial effects are negligible.

 

\section{Results}

\noindent\textbf{Polymer solution homogenizes flow above a threshold Weissenberg number, coinciding with the onset of elastic turbulence.} We use our stratified Hele-Shaw assembly to characterize the uneven partitioning of flow between strata at the macro-scale. First, we impose a small flow rate $Q=3$ mL/hr corresponding to $\mathrm{Wi}_I=1.4$---below the onset of elastic turbulence at $\mathrm{Wi}_I\approx2.6$ for homogeneous media \citep{Browne2021}. As is the case with Newtonian fluids, we observe preferential flow through the coarse stratum, indicated by the infiltrating dye front in the first panel of figure \ref{fig:1}B and in movie S1. The infiltration of dye at different rates through the strata produces two distinct steps in the breakthrough curve (dark green line in Figure \ref{fig:1}C): the first jump from $\tilde{C}\approx0$ to $0.4$ from {$0<\tilde{t}\lesssim3$} corresponds to fluid infiltration of the coarse stratum, and the second jump from $\tilde{C}\approx0.4$ to $0.8$ from {$3\lesssim\tilde{t}\lesssim6$} corresponds to infiltration of the fine stratum. This uneven partitioning of flow is also reflected in the difference between the magnitudes of the superficial velocities $U_C=130$ $\muup$m/s and $U_F=10$ $\muup$m/s in the coarse and fine strata, respectively, corresponding to a ratio of $U_F/U_C=0.075$. We observe similar behavior with our Newtonian control, which produces a similar ratio of $\left(U_F/U_C\right)_{0}=0.063$ even at a larger imposed flow rate $Q=35~\mathrm{mL/hr}$ (Movie S2). Hence, at low $\mathrm{Wi}_I$, polymer solutions recapitulate the uneven partitioning of flow across strata that is characteristic of Newtonian fluids.


Next, we repeat the same experiment as in figure \ref{fig:1}B at {a} larger flow rate of {$Q=25~\mathrm{mL/hr}$}---corresponding to a larger $\mathrm{Wi}_I=2.7$. Surprisingly, under these conditions, the partitioning of flow is markedly less uneven (second panel of figure \ref{fig:1}B, movie S3). These observations are reflected in the dye breakthrough curve, as well: the previously distinct jumps in the concentration $\tilde{C}$ begin to merge, as shown by comparing the light green and green lines in figure \ref{fig:1}C. Indeed, the ratio between the superficial velocities in the fine and coarse strata {$U_F/U_C=0.16$}, $\sim3\times$ larger than in the laminar baseline given by the Newtonian control and the low $\mathrm{Wi}_I=1.4$ solution tests. Therefore, to quantify this net improvement in flow homogenization, we normalize the velocity ratio by its Newtonian value, {$\tilde{U}_F/\tilde{U}_C\equiv\left(U_F/U_C\right)/\left(U_F/U_C\right)_{0}=2.6$}. This improvement in the flow homogenization is weaker at an even larger flow rate $Q=45~\mathrm{mL/hr}$ (corresponding to $\mathrm{Wi}_I=3.3$), as shown in the third panel of figure \ref{fig:1}B, the dark green line in figure \ref{fig:1}C, and in movie S4; the corresponding velocity ratio is $\tilde{U}_F/\tilde{U}_C=1.7$. Taken together, our observations demonstrate that high-molecular weight polymer additives can help mitigate uneven partitioning of flow in a stratified porous medium---but that this effect is optimized at intermediate $\mathrm{Wi}_I$.



Why does this flow homogenization arise? To shed light on the underlying physics, we use our ``continuous'' imaging protocol to directly image the flow at the pore scale within the stratified microfluidic assembly. At the intermediate $\mathrm{Wi}_I=2.7$---at which the flow homogenization is optimized---all pores observed in the fine stratum exhibit laminar flow that is steady over time (Movie S5; representative pore shown in the top left panel of figure \ref{fig:1}D). By contrast, 20\% of the pores observed in the coarse stratum exhibit strong spatial and temporal fluctuations in the flow (Figure \ref{fig:1}E). The fluid streamlines continually cross and vary over time, indicating the emergence of an elastic instability, as shown in Movie S5 and in the bottom left panel of figure \ref{fig:1}D for a representative pore. These random streamline fluctuations are similar to those observed for elastic turbulence \footnote{We note that the term ``elastic turbulence'' is sometimes used to refer to a particular manifestation of this unstable flow state at large Weissenberg numbers characterized by specific power-law decays of the power spectra of flow fluctuations. In this paper, we use the term to refer more generally to polymer elasticity-generated flow instabilities at low Re $\ll1$.} in a homogeneous medium \citep{Browne2021}; to highlight the regions of unstable flow, the figure also includes an overlay in red showing the standard deviation of the fluctuations in the fluorescent intensity over time. At the even larger $\mathrm{Wi}_I=3.3$---at which the improvement in flow homogenization is weaker---a larger fraction of pores in \textit{both} strata exhibit unstable flow (Movie S6; righthand panels of figure \ref{fig:1}D, figure \ref{fig:1}E). These results thus suggest that macroscopic flow homogenization is linked to the onset of elastic turbulence in the coarse stratum at sufficiently large $\mathrm{Wi}_I$, but is mitigated by the additional onset of elastic turbulence in the fine stratum at even larger $\mathrm{Wi}_I$.\\

\noindent\textbf{Flow fluctuations generated by elastic turbulence lead to an increase in the apparent viscosity. }
To quantitatively understand the link between pore-scale differences in this flow instability and macro-scale differences in superficial velocity between strata, we consider the resistance to flow in the distinct strata at different $\mathrm{Wi}_I$. In particular, we model the strata as parallel fluidic ``resistors''---that is, we treat each stratum as a homogeneous porous medium (e.g., coarse $C$ or fine $F$), with the two hydraulically connected only at the inlet and outlet with fully-developed flow in each. Because the time-averaged pressure drop $\langle\Delta P\rangle_t$ is equal across both strata, the imposed constant volumetric flow rate $Q$ must partition into the coarse and fine strata with flow rates $Q_C$ and $Q_F$, respectively, in proportion to their individual flow resistances via Darcy's law:
\begin{equation}
   \frac{\langle\Delta P\rangle_t}{L}=\frac{\eta_{\text{app},C}Q_C}{k_CA_C}
    =\frac{\eta_{\text{app},F}Q_F}{k_FA_F}.\label{eq:pressurebalance}
\end{equation} 
Following our previous study of elastic turbulence in a homogeneous porous medium \citep{Browne2021}, we combine macro-scale pressure drop measurements with pore-scale flow visualization to determine and validate a model for the $\eta_{\text{app},i}$ of each stratum in isolation. We then use this model to deduce the apparent viscosity and uneven partitioning of flow within a stratified medium. 

\begin{figure*}
    \centering
    \includegraphics[width=\textwidth]{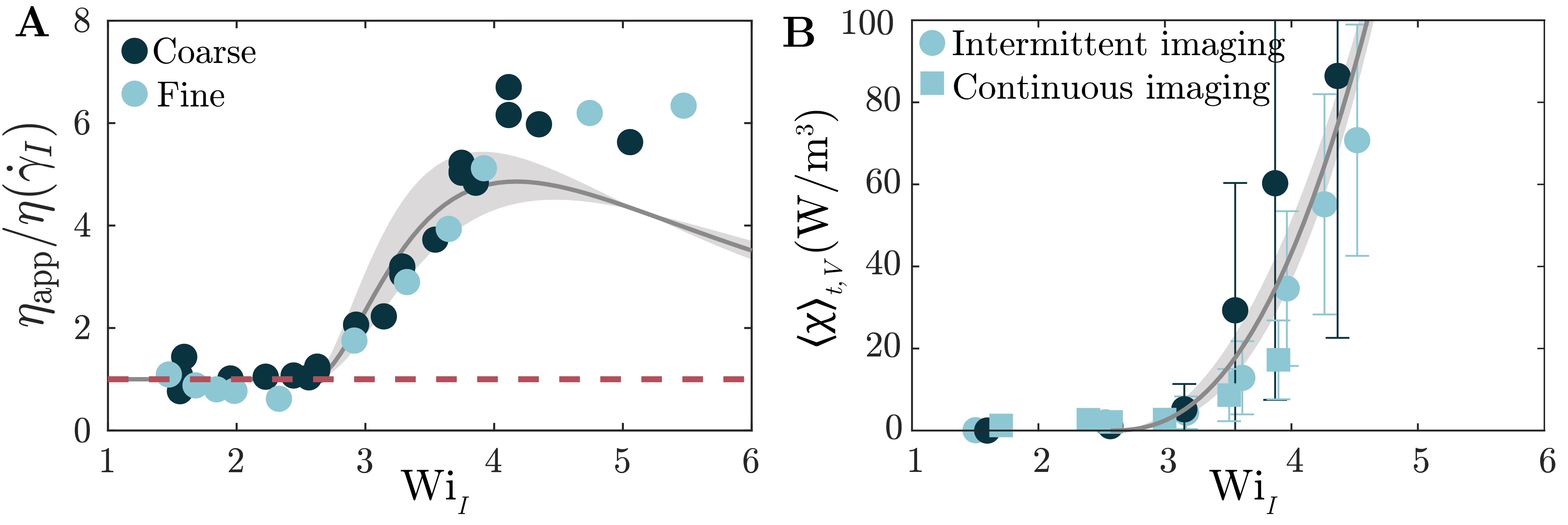}
    \caption{\textbf{Elastic turbulence produces a similar increase in the macroscopic flow resistance for homogeneous porous media of different permeabilities.} \textbf{A} Points show the apparent viscosity, normalized by the shear viscosity of the bulk solution, obtained using macroscopic pressure drop measurements. The apparent viscosity increases above a threshold $\mathrm{Wi}_I$ due to the onset of elastic turbulence. Measurements for two different homogeneous media with distinct bead sizes and permeabilities (different colors) show similar behavior. Grey line shows the predicted apparent viscosity using our power balance (equation~\ref{eq:appVisc}, neglecting strain history) and the measured power-law fit to $\langle\chi\rangle_{t,V}$ shown in \textbf{B}, with no fitting parameters; the uncertainty associated with the fit yields an uncertainty in this prediction, indicated by the shaded region. At the largest $\mathrm{Wi}_I$, the apparent viscosity eventually converges back to the shear viscosity, reflecting the increased relative influence of viscous dissipation from the base laminar flow. \textbf{B} Points show the rate of added viscous dissipation due to unstable flow fluctuations averaged over the medium, $\langle\chi\rangle_{t,V}$, measured from flow visualization. The dissipation sharply increases above the onset of elastic turbulence and is not sensitive to the bead size. Error bars represent one standard deviation between pores. We fit the data using an empirical power-law relationship $\sim(\mathrm{Wi}_I/\mathrm{Wi}_c-1)^{2.4}$ above the macroscopic threshold $\mathrm{Wi}_c=2.6$, shown by the grey line; the shaded region shows the error in the power-law fit.}
    \label{fig:2}
\end{figure*}

To do so, we measure the time-averaged pressure drop $\langle\Delta P\rangle_t$ at different volumetric flow rates $Q$ across each microfluidic assembly. We use Darcy's law to determine the corresponding $\eta_{\text{app}}$, which we plot as a function of $\mathrm{Wi}_I$ in figure \ref{fig:2}A; the data for the coarse medium are taken from \cite{Browne2021}. As expected, at small $\mathrm{Wi}_I\lesssim2.6$, the apparent viscosity $\eta_{\text{app}}$ is given by the bulk solution shear viscosity $\eta\left(\dot{\gamma}_I\right)$, indicated by the red dashed line. However, above a threshold $\mathrm{Wi}_c=2.6$, $\eta_{\text{app}}$ rises sharply, paralleling previous reports \citep{marshall1967flow,james1975laminar,durst1981,dursthaas1981,kauser,clarke2016}. Both the homogeneous coarse (dark blue circles) and fine (light blue circles) media exhibit a similar dependence of $\eta_{\text{app}}$ on $\mathrm{Wi}_I$---indicating that for our geometrically-similar packings, $\eta_{\text{app}}(\mathrm{Wi}_I)$ does not depend on grain size $d_p$.


To model this dependence of $\eta_{\text{app}}$ on $\mathrm{Wi}_I$, we directly image the pore-scale flow in each homogeneous microfluidic assembly with confocal microscopy {using our ``intermittent'' imaging protocol}. We previously reported these measurements solely for the homogeneous coarse medium \citep{Browne2021}; thus, we first summarize these results. At small $\mathrm{Wi}_I<2.6$, the flow is laminar in all pores. Above the threshold $\mathrm{Wi}_c=2.6$, the flow in some pores becomes unstable, exhibiting strong spatiotemporal fluctuations. At progressively larger $\mathrm{Wi}_I$, an increasing fraction of the pores becomes unstable. To directly compute the added viscous dissipation arising from these flow fluctuations, we measure the instantaneous 2D velocities $\textbf{u}$ using particle image velocimetry (PIV) \citep{thielicke2014pivlab}. Subtracting off the temporal mean in each pixel yields the velocity fluctuation $\textbf{u}'=\textbf{u}-\langle\textbf{u}\rangle_t$, from which we compute the fluctuating component of the strain rate tensor $\textbf{s}'=(\nabla\textbf{u}'+\nabla\textbf{u}'^{\mathrm{T}})/2$. The rate of added viscous dissipation per unit volume arising from these flow fluctuations is then given directly by $\langle\chi\rangle_t=\eta\langle\textbf{s}':\textbf{s}'\rangle_t$, which can be estimated from the measured 2D velocity field \citep{delafosse2011estimation,sharp2000dissipation}. 
As anticipated, the overall rate of added dissipation per unit volume $\langle\chi\rangle_{t,V}$ determined by averaging $\langle\chi\rangle_t$ across all imaged pores increases with $\mathrm{Wi}_I$ above the threshold $\mathrm{Wi}_c=2.6$ (Figure \ref{fig:2}B, dark blue circles) as a greater fraction of pores becomes unstable.

Next, we repeat this procedure in the homogeneous fine medium (Figure \ref{fig:2}B, light blue circles). Intriguingly, the overall rate of added dissipation per unit volume $\langle\chi\rangle_{t,V}$ does not significantly vary between media. Additionally measuring $\langle\chi\rangle_{t,V}$ using our ``continuous'' imaging protocol in the homogeneous fine medium further corroborates this agreement (Figure \ref{fig:2}B, light blue squares). We speculate that this collapse reflects that flow fluctuations do not have a characteristic length scale \citep{Browne2021}; further studies of the influence of confinement on $\langle\chi\rangle_{t,V}$ will be a useful direction for future work. Our data indicate that, for the experiments reported here, differences in grain size between homogeneous porous media are well-captured by $\mathrm{Wi}_I$. We therefore fit all the data by the single empirical relationship $\langle\chi\rangle_{t,V}=A_x(\mathrm{Wi}_I/\mathrm{Wi}_c-1)^{\alpha_x}$, with $A_x=176\pm1~\mathrm{W/m^3}$, $\alpha_x=2.4\pm0.3$, and $\mathrm{Wi}_c=2.6$, shown by the grey line in figure \ref{fig:2}B.



Finally, we follow our previous work \citep{Browne2021} to quantitatively link the pore-scale flow fluctuations generated by elastic turbulence to $\eta_{\text{app}}(\mathrm{Wi}_I)$. The power density balance for viscous-dominated flow relates the rate of work done by the fluid pressure $P$ to the rate of viscous energy dissipation per unit volume: $-\nabla\cdot P\textbf{u}=\boldsymbol{\tau}:\nabla\textbf{u}$, where $\tau$ and $\nabla\textbf{u}$ are the stress and velocity gradient tensors, respectively. Averaging this equation over time $t$ and the entire volume $V$ of a given porous medium, and decomposing the velocity field into the sum of a base temporal mean and an additional component due to velocity fluctuations, then yields:
\begin{equation}\label{eq:appVisc}
    \frac{\langle\Delta P\rangle_{t}}{\Delta L}\equiv\frac{\eta_{\text{app}} (Q/A)}{k}\approx{\underbrace{\frac{\eta(\dot{\gamma}_{I}) (Q/A)}{k}}
    _{\text{Darcy's law}}}\,
    +{\underbrace{\frac{\langle\chi\rangle_{t,V}}{(Q/A)}}_{\text{Fluctuations}}}+\left\{\rule{0cm}{.75cm}\parbox{3.5em}{\centering Strain\\history\\effects}\right\}\rule{0cm}{.75cm}.
\end{equation}
The first term on the right-hand side of Eq.~\ref{eq:appVisc} represents Darcy’s law for the base temporal mean of the flow. The second term reflects the added viscous dissipation by the solvent induced by the unstable flow fluctuations. The final term represents additional contributions arising from the full dependence of stress $\tau$ on polymer strain history in 3D~\citep{bird1987dynamics}, which is currently inaccessible in our experiments. However, our previous measurements in the homogeneous course medium \citep{Browne2021} indicate that this final term is relatively small for the range of $\mathrm{Wi}_I$ considered here, because the flow is quasi-steady and polymers do not accumulate appreciable Hencky strain over a duration of one polymer relaxation time $\lambda$. Therefore, for simplicity, we consider just the first two terms, which yields the grey line in figure \ref{fig:2}A; the shaded region indicates the uncertainty in this model arising from the empirical fit in figure \ref{fig:2}B. Our modeled $\eta_{\text{app}}(\mathrm{Wi}_I)$ thereby obtained from the pore-scale imaging shows excellent agreement with the $\eta_{\text{app}}$ obtained from the macro-scale pressure drop measurements (symbols) for both homogeneous media, without using any fitting parameters, for $\mathrm{Wi}_I\lesssim4$. The slight discrepancies at larger $\mathrm{Wi}_I$ suggest that strain history effects play a non-negligible role in this regime. Nevertheless, as a first approximation, we use the $\eta_{\text{app}}(\mathrm{Wi}_I)$ modeled using equation~\ref{eq:appVisc} (neglecting the last term describing strain history) to deduce the apparent viscosity $\eta_{\text{app},i}$ within each stratum in equation~\ref{eq:pressurebalance}.\\


\begin{figure*}
    \centering
    \includegraphics[width=\textwidth]{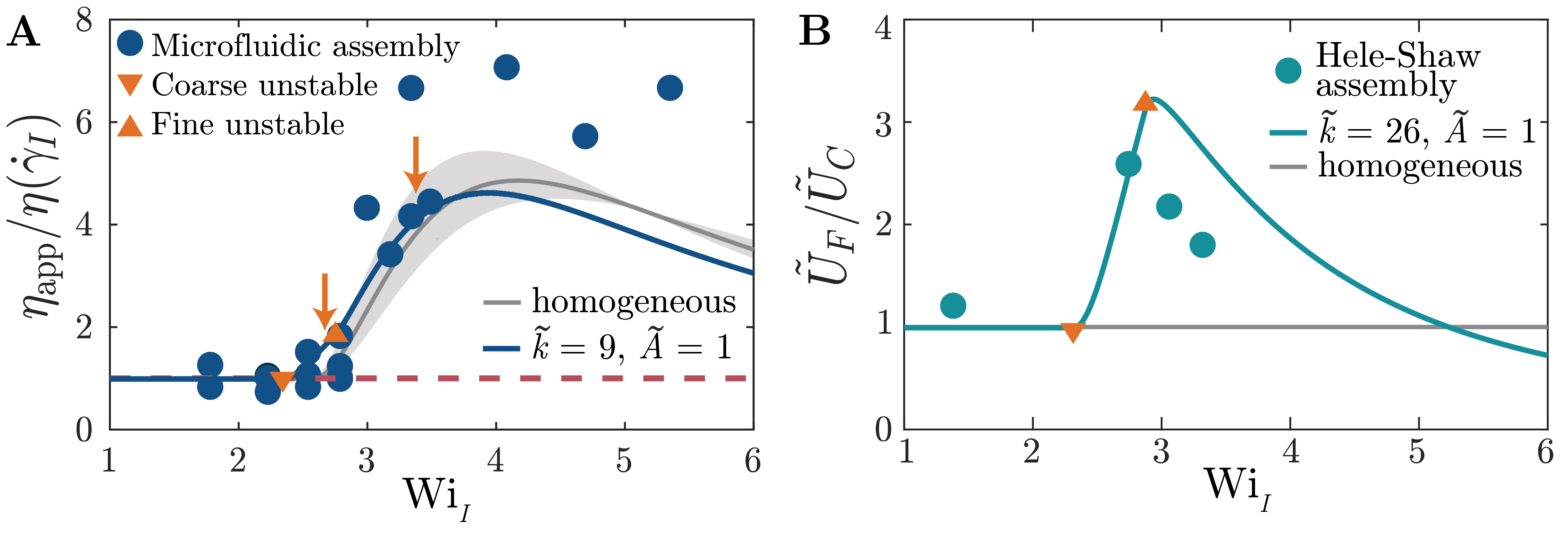}
    \caption{\textbf{Parallel-resistor model captures the key features of experimentally-measured apparent viscosity and uneven flow partitioning in stratified media.} \textbf{A}  Points show the normalized apparent viscosity measured for a stratified microfluidic assembly, indicating that it shows a similar increase above the onset of elastic turbulence. Blue line shows the predicted apparent viscosity using our parallel-resistor model with no fitting parameters. Grey line shows the corresponding prediction for a homogeneous medium. Left and right arrows show $\mathrm{Wi}_I=2.7$ and $3.3$, at which only the coarse stratum or both strata are unstable in figure~\ref{fig:1}D, respectively. The downward and upward triangles indicate the $\mathrm{Wi}_I$ at which each stratum becomes unstable. \textbf{B} Points show the ratio of superficial velocities in each stratum, normalized by the Newtonian value, measured for a stratified Hele-Shaw assembly; $\tilde{U}_{F}/\tilde{U}_C$ increases above the onset of elastic turbulence in the coarse stratum, indicating flow homogenization, and then decreases above the onset of elastic turbulence in the finer stratum as well, indicating that flow homogenization is mitigated. Teal line shows the prediction from our parallel-resistor model, which captures this non-monotonic behavior.}
    \label{fig:validation}
\end{figure*}

\noindent\textbf{Parallel-resistor model recapitulates experimental measurements of apparent viscosity and uneven flow partitioning.} We next incorporate our model for the apparent viscosity $\eta_{\text{app},i}(\mathrm{Wi}_I)$ in the parallel-resistor model of a stratified medium described previously. Specifically, for a given imposed total flow rate $Q$, which corresponds to a given $\mathrm{Wi}_I$, we numerically solve equations~\ref{eq:pressurebalance} and \ref{eq:appVisc} (neglecting the last term) along with mass conservation ($Q=Q_F+Q_C$) to obtain the apparent viscosity $\eta_{\text{app}}(\mathrm{Wi}_I)$ for the entire stratified system. 

To validate this approach, we first compute $\eta_{\text{app}}(\mathrm{Wi}_I)$ for the case of $\tilde{k}=9$ and $\tilde{A}=1$, which describes the stratified microfluidic assembly used in our experiments. Notably, the model shows a similar threshold $\mathrm{Wi}_c=2.6$ and overall shape of $\eta_{\text{app}}(\mathrm{Wi}_I)$ as in the homogeneous case, as shown by the blue line in figure~\ref{fig:validation}A---suggesting that stratification does not appreciably alter the macroscopic flow resistance. Indeed, we find good agreement between this model prediction and our experimentally-determined $\eta_{\text{app}}$, obtained from pressure drop measurements across the stratified microfluidic assembly, as shown by the blue circles in figure~\ref{fig:validation}A. 

This model also enables us to predict the onset of elastic turbulence in the different strata at different values of the macroscopic $\mathrm{Wi}_I$. As demonstrated by the experiments on homogeneous media (Figure~\ref{fig:2}), a given stratum becomes unstable when the \textit{local} Weissenberg number exceeds the threshold $\mathrm{Wi}_c=2.6$. However, because of the difference in the permeabilities of the strata, flow partitions unevenly across them, causing different strata to reach this threshold at different imposed macroscopic $\mathrm{Wi}_I$. For small $\mathrm{Wi}_I$, the flow is slower in the fine stratum, with the ratio of superficial velocities given by the Newtonian value $\left(U_F/U_C\right)_0=0.075$. As a result, the model predicts that the coarse stratum becomes unstable at a smaller value of the macroscopic $\mathrm{Wi}_{c,C}=2.3$ (downward triangles in figure~\ref{fig:validation}), and that the fine stratum becomes unstable at an even larger $\mathrm{Wi}_{c,F}=2.8$ (upward triangles). This prediction is in excellent agreement with our experimental pore-scale observations (Figure~\ref{fig:1}D--E) that at $\mathrm{Wi}_{I}=2.7$ (left arrow in figure~\ref{fig:validation}A), only the coarse stratum is unstable, while at a larger $\mathrm{Wi}_{I}=3.3$ (right arrow), both strata are unstable.

The model also reproduces and sheds light on the physics underlying the flow homogenization induced by elastic turbulence, as we observed experimentally in the stratified Hele-Shaw assembly (Figure~\ref{fig:1}B--C). For this case of $\tilde{k}=26$ and $\tilde{A}=1$, we use the model to compute the normalized ratio of superficial velocities $\tilde{U}_F/\tilde{U}_C\equiv\left(U_F/U_C\right)/\left(U_F/U_C\right)_0$ as a function of $\mathrm{Wi}_I$. The model prediction is shown by the line in figure~\ref{fig:validation}B. As expected, with increasing $\mathrm{Wi}_I$, the onset of elastic turbulence in the coarse stratum increases the resistance to flow in this stratum, redirecting fluid toward the fine stratum and thereby homogenizing the uneven flow across the entire medium---as indicated by the rapid increase in $\tilde{U}_F/\tilde{U}_C$ above $\mathrm{Wi}_{c,C}=2.3$ (downward triangle). However, this homogenization only arises in a window of flow rates: at even larger $\mathrm{Wi}_{I}>\mathrm{Wi}_{I,F}=2.8$ (upward triangle), $\tilde{U}_F/\tilde{U}_C$ peaks and continually decreases, reflecting the onset of elastic turbulence in the fine stratum as well. While we do not expect perfect quantitative agreement with the experiments, given the assumptions and approximations made in our model, the experimental measurements show similar behavior: as shown by the circles in figure~\ref{fig:validation}B, $\tilde{U}_F/\tilde{U}_C$ initially rises for $\mathrm{Wi}_I>\mathrm{Wi}_{c,C}=2.3$, and then continues to decrease as $\mathrm{Wi}_I$ exceeds $\mathrm{Wi}_{c,F}=2.8$.

Thus, despite its simplicity, the parallel-resistor model of a stratified medium (Equation~\ref{eq:pressurebalance}) that explicitly incorporates the increase in flow resistance generated by elastic turbulence in each stratum (Equation~\ref{eq:appVisc}) captures our key experimental findings: (i) the form of the macroscopic $\eta_{\text{app}}(\mathrm{Wi}_I)$ describing the entire medium, (ii) the differential onset of elastic turbulence in the different strata at varying $\mathrm{Wi}_I$, and (iii) the corresponding window of $\mathrm{Wi}_I$ within which the uneven flow across strata is homogenized. Having thereby validated the model, we next use it to further examine how elastic turbulence may homogenize fluid transport in stratified porous media having a broader range of permeability and area ratios, $\tilde{k}\equiv k_{C}/k_{F}$ and $\tilde{A}\equiv A_{C}/A_{F}$, respectively, than currently accessible in the experiments.\\

\begin{figure*}
    \centering
    \includegraphics[width=\textwidth]{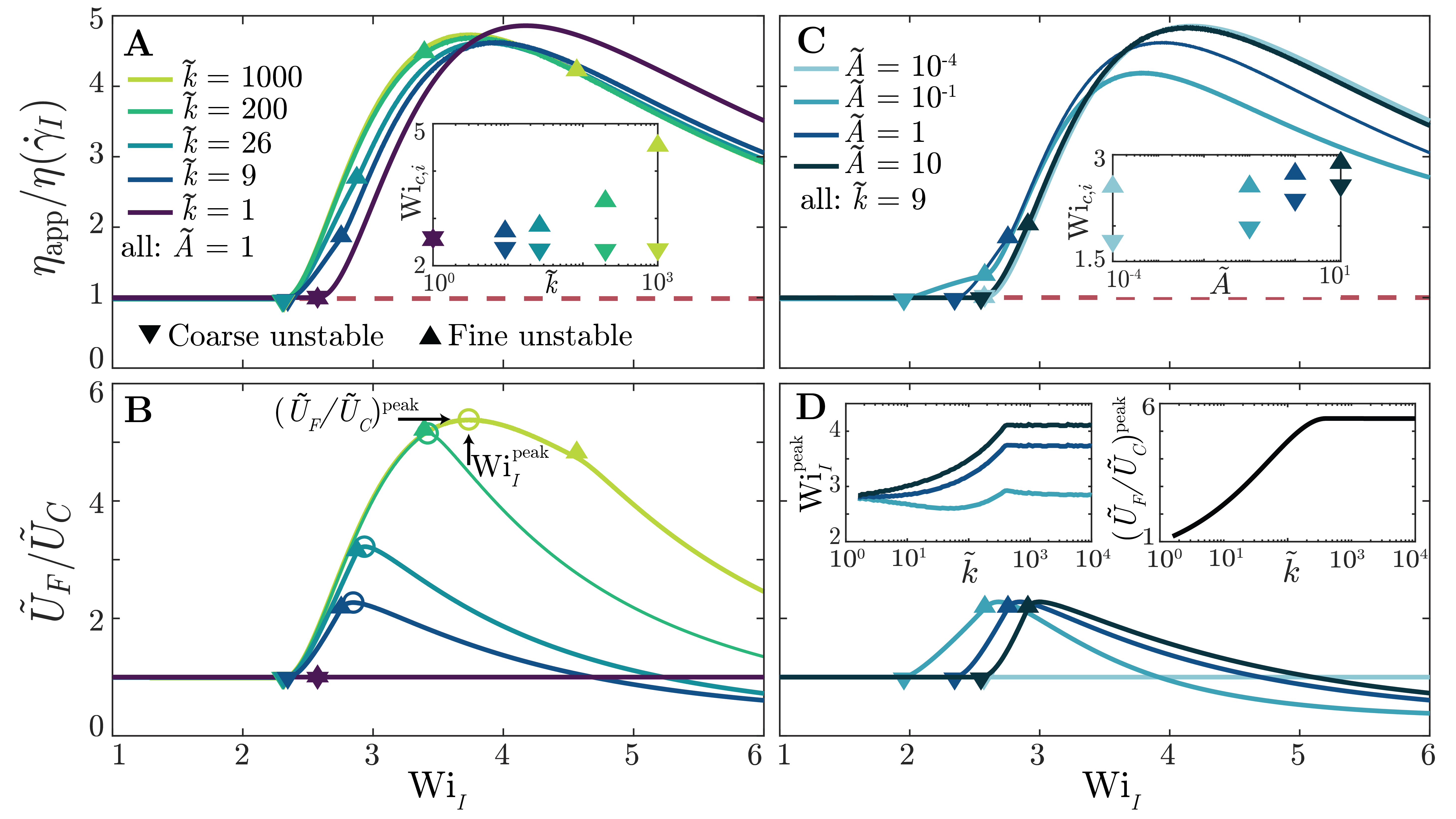}
    \caption{\textbf{Geometry dependence of the apparent viscosity and uneven flow partitioning in a stratified medium, as predicted by our parallel-resistor model.} \textbf{A--B} Different colors show the predictions of the parallel-resistor model for stratified media with varying ratios of the strata permeabilities, $\tilde{k}$, holding the area ratio fixed at $\tilde{A}=1$. The apparent viscosity (\textbf{A}) only shifts slightly to smaller $\mathrm{Wi}_{I}$ with increasing $\tilde{k}$, eventually converging for $\tilde{k}\gg100$. The extent of flow homogenization generated by elastic turbulence, quantified by the ratio of superficial velocities (\textbf{B}), does increase with increasing $\tilde{k}$. Optimal flow homogenization is indicated by the open circles at $\mathrm{Wi}_I=\mathrm{Wi}_I^\text{peak}$ with a velocity ratio $\left(\tilde{U}_F/\tilde{U}_C\right)^\text{peak}$. Inset to \textbf{A} shows the critical $\mathrm{Wi}_{I}$ at which each stratum becomes unstable; the window between the two values increases with increasing $\tilde{k}$. \textbf{C--D} Similar results to \textbf{A--B}, but for stratified media with varying strata area ratios, $\tilde{A}$, holding the permeability ratio fixed at $\tilde{k}=9$. Inset to \textbf{C} shows the critical $\mathrm{Wi}_{I}$ at which each stratum becomes unstable; the window between the two values decreases with increasing $\tilde{A}$. Insets to \textbf{D} show the variation of the optimal $\mathrm{Wi}_I^\text{peak}$ and $\left(\tilde{U}_F/\tilde{U}_C\right)^\text{peak}$ with $\tilde{k}$, for different $\tilde{A}$. The data for different $\tilde{A}$ trivially collapse due to the definition of the superficial velocity.}
    \label{fig:4}
\end{figure*}

\noindent\textbf{Geometry-dependence of flow homogenization.} How do the onset of and extent of homogenization imparted by elastic turbulence depend on the geometric characteristics of a stratified porous medium? To address this question, we use our model to probe how the overall apparent viscosity $\eta_{\text{app}}(\mathrm{Wi}_I)$ and the flow velocity ratio $\tilde{U}_F/\tilde{U}_C(\mathrm{Wi}_I)$ depend on $\tilde{k}$ and $\tilde{A}$.

The measurements shown in Figure~\ref{fig:validation} indicate that, despite the structural heterogeneity and uneven partitioning of the flow in a stratified medium, $\eta_{\text{app}}(\mathrm{Wi}_I)$ is not strongly sensitive to stratification; instead, it follows a similar trend to that of a homogeneous medium ($\tilde{k}=1$). The model further supports this finding; with increasing $\tilde{k}$ (fixing $\tilde{A}=1$), the profile of $\eta_{\text{app}}(\mathrm{Wi}_I)$ shifts ever so slightly to smaller $\mathrm{Wi}_I$, eventually converging to the same final profile for $\tilde{k}\gg100$, as shown in figure~\ref{fig:4}A. 

\begin{figure*}
    \centering
    \includegraphics[width=\textwidth]{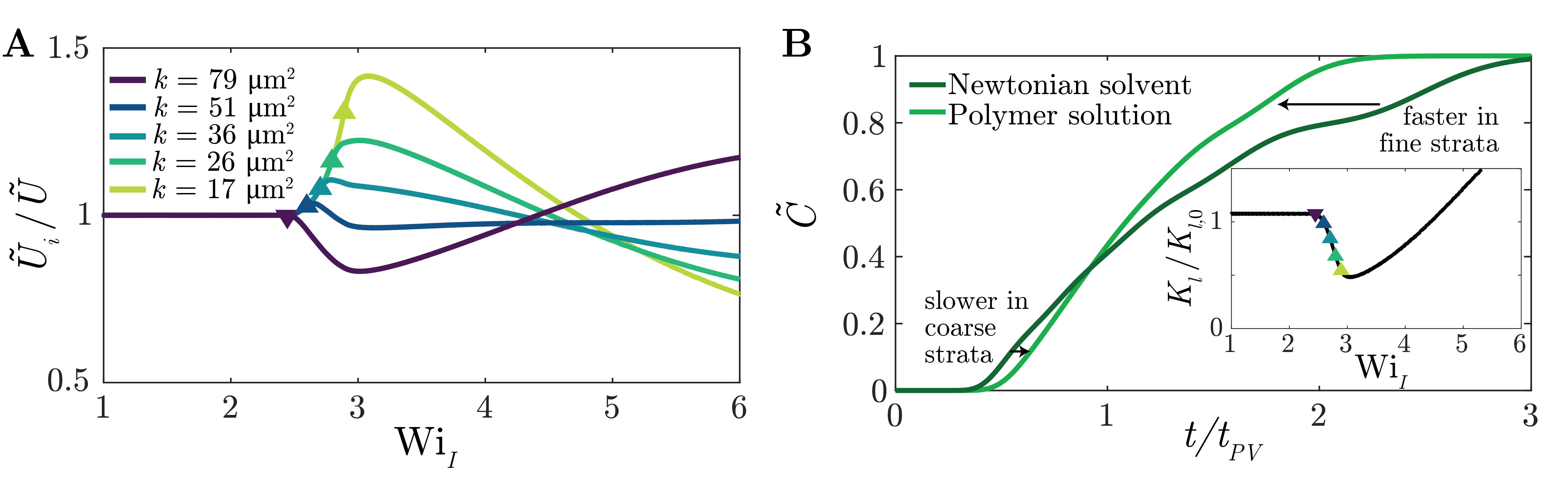}
    \caption{\textbf{Model predictions for a porous medium with five distinct strata}. \textbf{A} Different colors show the predicted ratio between the superficial velocity in each stratum and the macroscopic superficial velocity, normalized by the value of this ratio for a Newtonian polymer-free solvent. The coarsest stratum (dark purple) becomes unstable at the smallest $\mathrm{Wi}_I$, following by the next coarsest (dark blue) and so on---causing flow to be redirected to the finer strata and the uneven flow across different strata to be homogenized. at even larger $\mathrm{Wi}_I$, all the strata become unstable and the resulting flow homogenization is mitigated. \textbf{B} Predicted breakthrough curves for the polymer solution at $\mathrm{Wi}_I=3.2$ (light green) as well as the Newtonian polymer-free solvent at the same flow rate (dark green). At this intermediate $\mathrm{Wi}_I$, elastic turbulence homogenizes the uneven flow across strata; as a result, rapid breakthrough in the coarsest strata is slowed (left arrow), and slow breakthrough in the finest strata is hastened (right arrow), smoothing the overall breakthrough curve. Inset shows the macroscopic effective longitudinal dispersivity, normalized by its value for the Newtonian polymer-free solvent at the same volumetric flow rate. $K_l$ and $K_{l,0}$ differ slightly at low $\mathrm{Wi}_I$ because of the modest shear-thinning in the polymer solution, which increases the uneven partitioning of flow uniformly before the onset of unstable flow. For a window of $2.4\lesssim\mathrm{Wi}_I\gtrsim4.5$, the normalized dispersivity is smaller than one, indicating more uniform scalar transport due to the homogenized flow resulting from elastic turbulence.}
    \label{fig:5}
\end{figure*}

However, the onset of elastic turbulence in the different strata does vary with increasing $\tilde{k}$ (inset of figure~\ref{fig:4}A): $\mathrm{Wi}_{c,C}$ correspondingly shifts to slightly smaller $\mathrm{Wi}_I$, while $\mathrm{Wi}_{c,F}$ progressively shifts to larger $\mathrm{Wi}_I$, reflecting the increasingly uneven partitioning of the flow imparted by increasing permeability differences. As a result, the strength of the flow homogenization generated by elastic turbulence, as well as the window of $\mathrm{Wi}_I$ at which it occurs, increases with $\tilde{k}$ (Figure~\ref{fig:4}B). This phenomenon is optimized at the peak position indicated by the open circles, which occur at $\mathrm{Wi}_I=\mathrm{Wi}_I^\text{peak}$ with a flow velocity ratio $\left(\tilde{U}_F/\tilde{U}_C\right)^\text{peak}$. We therefore summarize our results by plotting both quantities as a function of $\tilde{k}$ (dark blue lines, insets to figure~\ref{fig:4}D). Again, both increase 
until $\tilde{k}\approx400$. For even larger $\tilde{k}$, $\mathrm{Wi}_I^\text{peak}$ plateaus at $\approx3.7$, while $\left(\tilde{U}_F/\tilde{U}_C\right)^\text{peak}$ plateaus at $\approx5.5$. This behavior reflects the non-monotonic nature of our model for $\eta_{\text{app},i}(\mathrm{Wi}_I)$; at such large permeability ratios, the coarse stratum reaches its maximal value of $\eta_{\text{app},C}$ at $\mathrm{Wi}_I<\mathrm{Wi}_{c,F}$, maximizing the extent of flow redirection to the fine stratum generated by elastic turbulence in the coarse stratum. These physics are also reflected in the values of $\mathrm{Wi}_I^\text{peak}$ and $\mathrm{Wi}_{c,F}$ (open circles and filled upward triangles in figure~\ref{fig:4}B, respectively); while the two match for small $\tilde{k}$, $\mathrm{Wi}_I^\text{peak}$ becomes noticeably smaller than $\mathrm{Wi}_{c,F}$ for $\tilde{k}\gtrsim400$.


Similar results arise with varying $\tilde{A}$ (fixing $\tilde{k}=9$), as shown in figure~\ref{fig:4}C--D. Here, $\tilde{A}<1$ and $\tilde{A}>1$ describe the case in which a greater fraction of the medium cross-section is occupied by the fine or coarse stratum, respectively; the limits of $\tilde{A}\rightarrow0$ and $\rightarrow\infty$ therefore represent a non-stratified homogeneous medium. While stratification again does not strongly alter $\eta_{\text{app}}(\mathrm{Wi}_I)$, we find that $\mathrm{Wi}_{c,C}$, $\mathrm{Wi}_{c,F}$, and $\mathrm{Wi}_I^\text{peak}$ increase with $\tilde{A}$. Furthermore, $\left(\tilde{U}_F/\tilde{U}_C\right)^\text{peak}$ does not depend on $\tilde{A}$, since the superficial velocity incorporates cross-sectional area by definition. Taken together, these results provide quantitative guidelines by which the macroscopic flow resistance, as well as the onset and extent of flow homogenization, can be predicted for a porous medium with two parallel strata of a given geometry. \\

\noindent\textbf{Extending the model to porous media with even more strata.} As a final demonstration of the utility of our approach, we extend it to the case of a porous medium with $n$ parallel strata, each indexed by $i$. To do so, we again maintain the same pressure drop across all the different strata (Equation~\ref{eq:pressurebalance}), with the apparent viscosity $\eta_{\text{app},i}$ in each given by equation~\ref{eq:appVisc}, and numerically solve these $n-1$ equations constrained by mass conservation, $Q=\Sigma_{i=1}^{n} Q_i$.

As an illustrative example, we consider $n=5$ with the different stratum permeabilities chosen from a log-normal distribution, as is often the case in natural settings~\citep{freeze1975stochastic}: $k_i\in\{79,51,36,26,17\}~\mathrm{\muup m}^2$. To characterize the flow redirection between strata at varying overall $\mathrm{Wi}_I$, we focus on the ratio of the superficial velocity $U_i$ in each stratum and the macroscopic superficial velocity $U\equiv Q/A$, normalized by the value of this ratio for a Newtonian fluid: $\tilde{U}_{i}/\tilde{U}\equiv \left(U_{i}/U\right)/\left(U_{i}/U\right)_0$. Hence, larger (smaller) values of $\tilde{U}_{i}/\tilde{U}$ indicate that fluid is being redirected to (from) a given stratum $i$. Consistent with our previous results, the coarsest stratum becomes unstable at the smallest $\mathrm{Wi}_I$ (dark purple line in figure~\ref{fig:5}A), redirecting fluid to the other strata---as indicated by the reduction in $\tilde{U}_{i}/\tilde{U}$ for $k_i=79~\mathrm{\muup m}^2$ as $\mathrm{Wi}_I$ increases above $\approx2.4$, and the concomitant increase in $\tilde{U}_{i}/\tilde{U}$ for the other strata (blue to light green lines). Each progressively finer stratum then becomes unstable at progressively larger $\mathrm{Wi}_I$, as indicated by the upward triangles, redirecting fluid from it to the other strata. Thus, as with the case of $n=2$ examined previously, the flow homogenization generated by elastic turbulence arises only in a window of $\mathrm{Wi}_I$.

As a final illustration of this point, we compute the corresponding breakthrough curve of a passive scalar, $\tilde{C}(t)$, given that such curves are commonly used to characterize transport in porous media for a broad range of applications. To do so, for a given stratum $i$ with $U_i$ determined from our parallel-resistor model, we use the foundational model of \cite{perkins1963review} as an example to compute $C_i(t)=0.5\left[1-\mathrm{erf}\left(\frac{1-t/t_{PV}}{2\sqrt{K_{l,i}/U_iL}\sqrt{t/t_{PV}}}\right)\right]$. This expression explicitly incorporates the dispersion of a passive scalar being advected by the flow via the longitudinal dispersivity $K_{l,i}$, which depends on the scalar diffusion coefficient $D$, the stratum tortuosity $\tau$, and the P\'eclet number characterizing scalar transport in a pore $\mathrm{Pe}=U_i d_{p,i}/D$; in particular, $K_{l,i}=D(1/\tau+0.5\mathrm{Pe}^{1.2})$ when $\mathrm{Pe}<605$ and $K_{l,i}=D(1/\tau+1.8\mathrm{Pe})$ when $\mathrm{Pe}>605$ \citep{woods2015flow}. The overall breakthrough curve is then given by $C(t)=\sum_{i}^{n}C_i(t)A_i/A$, which we normalize by its maximal value at $t\rightarrow\infty$ to obtain $\tilde{C}(t)$. For this illustrative example, we use values characteristic of small molecule solutes in natural porous media: $D=10^{-6}~\mathrm{cm}^2/\mathrm{s}$, $\tau=2$~\citep{datta2013a}, and estimate $d_{p,i}$ from the stratum permeability using the Kozeny-Carman relation~\citep{philipse1993liquid}.

The resulting breakthrough curves $\tilde{C}(t)$ are shown in figure~\ref{fig:5}B for a fixed flow rate, chosen such that $\mathrm{Wi}_I=3.2$ for our polymer solution---just above the onset of elastic turbulence in the finest stratum, at which we expect flow homogenization to be nearly optimized (Figure~\ref{fig:5}A). For the case of the polymer-free Newtonian solvent, the flow partitions unevenly across the strata, leading to highly heterogeneous scalar breakthrough. As shown by the dark green line, coarser strata are infiltrated rapidly, leading to the rise in $\tilde{C}(t)$ at $t/t_{PV}\approx0.4$. However, the considerably smaller flow speeds in the bypassed finer strata give rise to far slower breakthrough, leading to the subsequent jumps in $\tilde{C}(t)$ at longer times; as a result, 90\% of scalar breakthrough only occurs after $t/t_{PV}=2.5$ has elapsed. The polymer solution exhibits strikingly different behavior: the breakthrough curve shown by the light green line is noticeably smoother, reflecting the flow homogenization imparted by elastic turbulence. In this case, unstable flow hinders rapid infiltration in the coarser strata (right-pointing arrow at $t/t_{PV}\approx0.6$), instead redirecting fluid to the finer strata (left-pointing arrow at $t/t_{PV}\approx2$); as a result, 90\% of scalar breakthrough occurs $\approx1.4\times$ faster, at $t/t_{PV}=1.8$. 

This improvement in scalar breakthrough can also be described using an effective, macroscopic, stratum-homogenized longitudinal dispersivity $K_{l}$. Despite the complex shapes of breakthrough curves that commonly arise for stratified porous media due to uneven flow partitioning (e.g., dark green line in figure~\ref{fig:5}B), a standard practice is to fit the entire breakthrough curve to a single error function \citep{lake1981taylor} and thereby extract $K_{l}$. The dispersitivy thereby determined from our computed breakthrough curves is shown in the inset to figure~\ref{fig:5}B for a broad range of $\mathrm{Wi}_I$. At small $\mathrm{Wi}_I$, $K_{l}$ matches that of a polymer-free Newtonian solvent $K_{l,0}$ at the same volumetric flow rate. Above $\mathrm{Wi}_I\approx2.4$, at which the coarsest stratum becomes unstable, $K_{l}$ drops relative to the Newtonian value---indicating more uniform scalar breakthrough due to flow homogenization. The effective dispersivity continues to decrease as an increasing number of strata become unstable, further homogenizing the flow and causing scalar breakthrough to become more uniform. The effective dispersity is ultimately minimized at the optimal $\mathrm{Wi}_I\approx3.2$. Increasing $\mathrm{Wi}_I$ further causes $K_{l}/K_{l,0}$ to then increase, eventually reaching 1 at $\mathrm{Wi}_I\approx4.5$---again reflecting the fact that the flow homogenization generated by elastic turbulence arises in the window of $2.4\lesssim\mathrm{Wi}_I\lesssim4.5$.

\section{Conclusions}
The work described here provides the first, to our knowledge, characterization of elastic turbulence in stratified porous media. Our experiments combining flow visualization with pressure drop measurements revealed that elastic turbulence arises at different flow rates, corresponding to different $\mathrm{Wi}_I$, in different strata. Uneven partitioning of flow into the higher-permeability strata causes them to become unstable at smaller $\mathrm{Wi}_I$---redirecting the flow towards the lower-permeability strata, thereby helping to homogenize the flow across the entire medium. At even larger $\mathrm{Wi}_I$, the lower-permeability strata become unstable as well, suppressing this flow redirection---leading to a window of flow rates at which this homogenization arises. 

We elucidated the physics underlying this behavior using a minimal parallel-resistor model of a stratified medium that explicitly incorporates the increase in flow resistance generated by elastic turbulence in each stratum. Despite the simplicity of the model, it captures the macroscopic resistance to flow through the entire medium, the differential onset of elastic turbulence in the different strata at varying $\mathrm{Wi}_I$, and the corresponding window of $\mathrm{Wi}_I$ within which the uneven flow across strata is homogenized, as found in the experiments. Taken together, our work thus establishes a new approach to homogenizing fluid and passive scalar transport in stratified porous media---a critical requirement in many environmental, industrial, and energy processes.

This study focused on a single polymer solution formulation as an illustrative example. However, the threshold $\mathrm{Wi}_c$ at which elastic turbulence arises, and the corresponding excess flow resistance $\langle\chi\rangle_{t,V}$, likely depend on the solution rheology (through e.g., polymer concentration, molecular weight, and solvent composition). The relative importance of the full polymer strain history in 3D, neglected here for simplicity, may also play a non-negligible role for different formulations and at large $\mathrm{Wi}_I$; indeed, while we use the specific functional form of $\eta_{\text{app}}$ given by equation~\ref{eq:appVisc}, it is unclear how far this model can be extrapolated past $\mathrm{Wi}_I\gtrsim4$. Incorporating these additional complexities into our analysis will be an important direction for future work.

Nevertheless, the theoretical framework established here provides a way to develop quantitative guidelines for the design of polymeric solutions and fluid injection strategies, given a stratified porous medium of a particular geometry. We therefore anticipate it will find use in diverse applications---particularly those that seek to balance the competing demands of minimizing the macroscopic resistance to flow (quantified by $\eta_{\text{app}}$) and maximizing flow homogenization (quantified by $\tilde{U}_i$). Indeed, accomplishing this balance is a critical challenge in subsurface processes such as pump-and-treat remediation of groundwater, in situ remediation of groundwater aquifers using injected chemical agents, enhanced oil recovery, and maximizing fluid-solid contact for heat transfer in geothermal energy extraction---for which uneven flow across strata is highly undesirable. Moreover, similar flows also play key roles in determining separation performance in filtration and chromatography, and improving heat and mass transfer in microfluidic devices. Thus, by deepening fundamental understanding of how elastic turbulence can be harnessed to homogenize flow in stratified media, we expect our results to inform a broader range of applications.\\


\noindent\textbf{Acknowledgements. }{It is our pleasure to acknowledge the Stone Lab for use of the rheometer. This material is based upon work supported by the National Science Foundation Graduate Research Fellowship Program (to C.A.B.) under Grant No. DGE-1656466, as well as partial support from Princeton University's Materials Research Science and Engineering Center under NSF Grant No. DMR-2011750. Any opinions, findings, and conclusions or recommendations expressed in this material are those of the authors and do not necessarily reflect the views of the National Science Foundation. C.A.B. was also supported in part by a Mary and Randall Hack Graduate Award from the High Meadows Environmental Institute and a Wallace Memorial Honorific Fellowship from the Graduate School of Princeton University.}\\

\noindent\textbf{Author contributions. }{C.A.B. and S.S.D. designed the experiments; C.A.B. performed all experiments; C.A.B. and S.S.D. designed the theoretical model; C.A.B. and R.B.H. performed all theoretical analysis and numerical simulations of 2-strata media; C.A.B. and C.W.Z. performed all theoretical analysis and numerical simulations of $n$-strata media; C.A.B. and S.S.D. analyzed all data, discussed the results and implications, and wrote the manuscript; S.S.D. designed and supervised the overall project.}


\end{document}